\documentclass[amsmath,amssymb,12pt]{article}

\usepackage{graphicx}
\usepackage{amssymb}
\usepackage{epstopdf}
\usepackage{pdfsync}

\usepackage{color}

\usepackage{graphicx}
\usepackage{amsmath}

\input{tcilatex}

\usepackage{amsfonts}
\usepackage{mathrsfs}

\usepackage{makeidx}

\makeindex

\begin{document}
 
  \title{Information States in Control Theory: \\ From Classical to Quantum
 \thanks{This research was supported by the Australian Research Council Centre of Excellence for Quantum Computation and Communication Technology (project number CE110001027),  and by US Air Force Office of Scientific Research Grant FA2386-09-1-4089.
 Dedicated to Bill Helton. Publication details: Harry Dym, Mauricio C. de Oliveira, Mihai Putinar (Eds.)
 Mathematical Methods in Systems, Optimization, and Control,
 Operator Theory: Advances and Applications Volume 222, 2012, pp 233-246.
 }
\author{M.R.~James\thanks{ARC Centre for Quantum Computation and Communication Technology, Research School of Engineering, Australian
National University, Canberra, ACT 0200, Australia (e-mail: 
Matthew.James@anu.edu.au)}
  }}
 \date{2012}

\maketitle

\begin{abstract}
This paper is concerned with the concept of {\em information state} and its use in  optimal feedback control of classical and quantum systems. The use of information states for {\em measurement} feedback problems is summarized. Generalization to   fully quantum coherent feedback  control problems is considered.
\end{abstract}

\section{Introduction}
\label{sec:intro}

This paper is dedicated to Bill Helton, with whom I had the honor and pleasure of collaborating   in the topic area of nonlinear $H^\infty$ control theory, \cite{HJ99}. We developed in some detail the application of information state methods to the nonlinear $H^\infty$ control problem,  \cite{JBE94,JB95,BB95}. In this paper I review the information state concept for classical output feedback optimal control problems, and then discuss extensions of this concept to quantum feedback control problems,  \cite{J04,J05,JNP08}.

\begin{figure}[h]
\centering
\includegraphics[width=12cm]{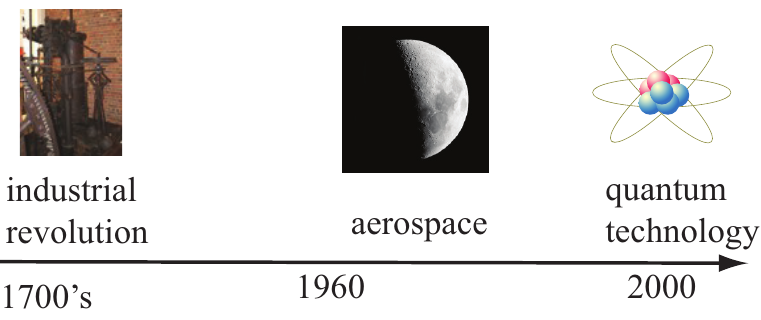}

\caption{Feedback control timeline.}
\label{fig:timeline1} 
\end{figure}

{\em Feedback} is  the most  important  idea in control engineering, and feedback 
 is a critical enabler for technological development, Figure \ref{fig:timeline1}.
From its origins in steam engine governors, through applications in electronics, aerospace, robotics, telecommunications and elsewhere, the use of feedback control has been essential in shaping our modern world. 
In the 20th century, quantum technology, through semiconductor physics and microchips, made possible the information age.  New developments in quantum technology, which include quantum information and computing, precise metrology, atom lasers, and quantum electromechanical systems, further exploit quantum phenomena and hold significant promise for the future.

\begin{figure}[h]
\centering
\includegraphics[width=10cm]{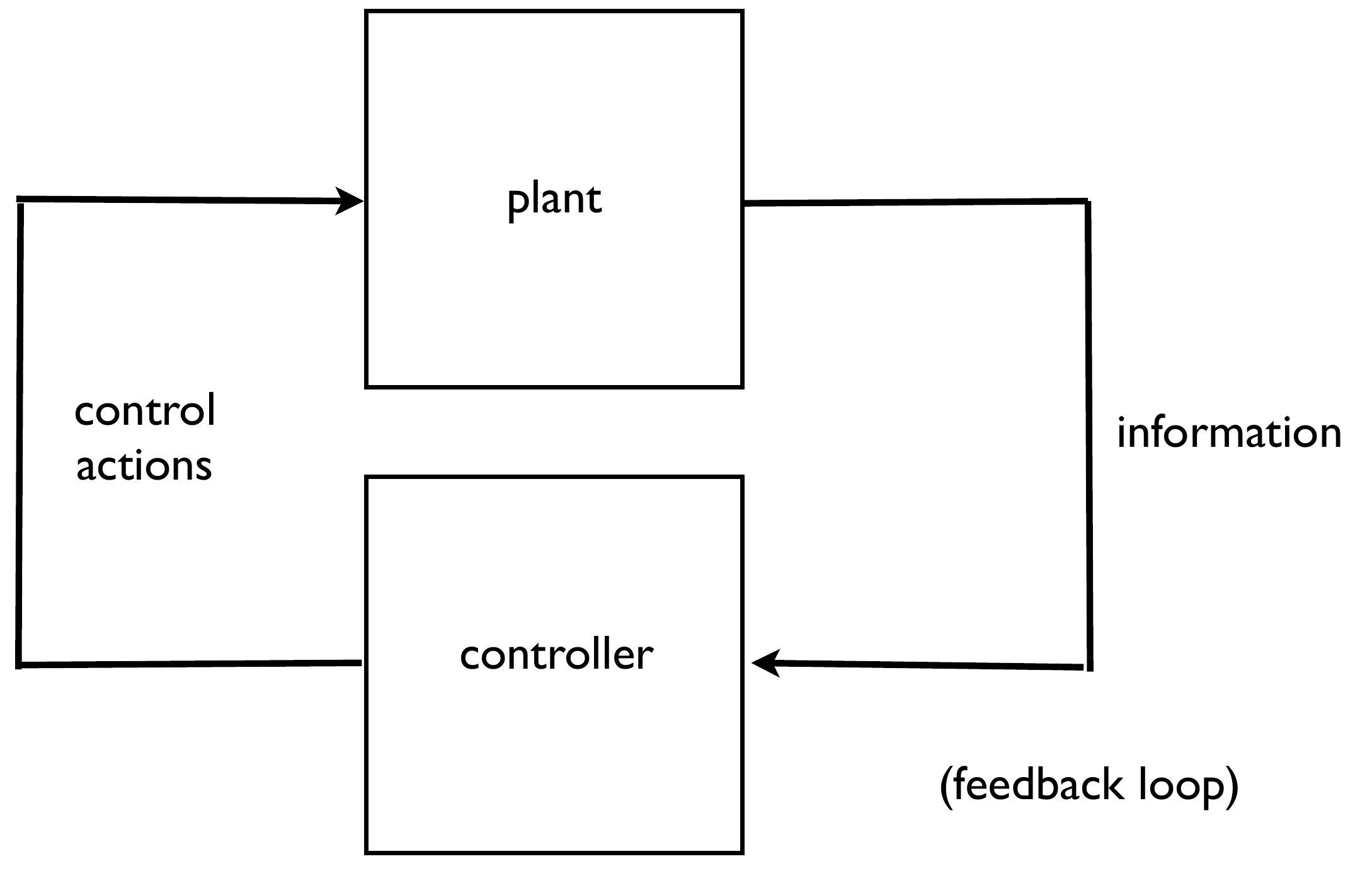}

\caption{Information flow in a feedback loop.}
\label{fig:feedback1} 
\end{figure}

{\em Optimization}  is  basic to many fields and is widely  used to design control systems.
 Optimization based control system design requires specification  of
(i)  the  {\em objective} of the control system, and (ii)  the {\em information} available to the control system. 
In a feedback system, Figure \ref{fig:feedback1},  control actions are determined on the basis of information gained as the system operates.
A key issue is how to represent information in a feedback loop. The concept of {\em information state} was introduced for this purpose, \cite{KV86}.
An  information state is a statistic\footnote{In statistics, a  {\em statistic}  is a measure of some attribute of a data sample.}  that takes into account the performance objective in a feedback loop.

 In quantum science and technology, the extraction of information about a system, and the use of this information for estimation and control, is a topic of fundamental importance.   The postulates of quantum mechanics specify the random nature of quantum measurements, and over a period of decades quantum measurement theory has led to a well developed theory of quantum conditional expectation and quantum filtering, \cite{VPB92,VPB92a,HC93,BHJ07,WM10}. Quantum filtering theory may be used as a framework for {\em measurement feedback} optimal control of quantum systems, and we summarize how this is done in Section \ref{sec:quantum}. In particular, we highlight the role of information states in this context. However, quantum measurement necessarily involves the loss of quantum information, which may not be desirable.  Fortunately, 
  feedback in quantum systems need not involve measurement. 
   In fully quantum coherent feedback, the physical system being controlled, as well as the device used for the controller, are quantum systems. 
 For instance, optical beams may be used to interconnect quantum devices and enable the transmission of quantum information from one system to another, thereby serving as \lq\lq{quantum wires}\rq\rq. To my knowledge,  to date there has been no extension of information states to fully quantum coherent  feedback optimal control, although it has been a topic of discussion. Instead, direct methods have been employed for special situations, \cite{JNP08,NJP09}. One of the key obstacles that makes optimal fully quantum coherent feedback control challenging is the general difficulties of conditioning onto non-commuting physical observables, a difficulty of fundamentally quantum mechanical origin (conditioning works successfully when measurements are used as then commuting observables are involved). 
  Section \ref{sec:quantum-coherent} discusses a possible means for 
abstracting the notion of information state may provide a suitable means for approaching the solution of optimal fully quantum feedback control problems in the context of a concrete example.

\section{Classical Output Feedback Optimal Control}
\label{sec:classical}

In many situations, information available to the controller is often partial, and subject to noise. 
In this section we  look at a standard scenario using stochastic models, and show how information states can be found for two types of performance criteria.

\begin{figure}[h]
\centering
\includegraphics[width=10cm]{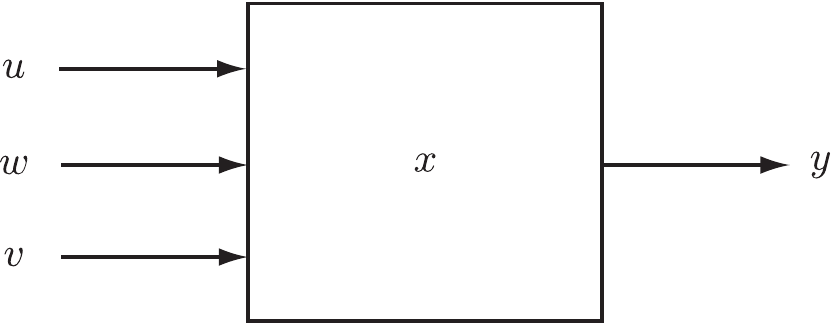}

\caption{A partially observed stochastic system with control input $u$ and observed output $y$. The internal state $x$ is not directly accessible.}
\label{fig:system1} 
\end{figure}

Consider the following
 Ito stochastic differential equation model 
\begin{eqnarray}
d x &=& f(x,u) dt + g(x) dw
\\
dy &=& h(x) dt + dv
\end{eqnarray}
where (i) 
$u$ is the control input signal, (ii)
$y$ is the observed output signal, (iii)
$x$ is a vector of internal state variables, and (iv)
$w$ and $v$ are independent standard Wiener processes.
 Note that $x(t)$ is a Markov process (given $u$) with generator
 $$
\mathcal{L}^u(\phi) = f(\cdot,u) \phi' + \frac{1}{2} g^2 \phi^{\prime\prime}
$$
The system is shown schematically in Figure \ref{fig:system1}

\begin{figure}[h]
\centering
\includegraphics[width=10cm]{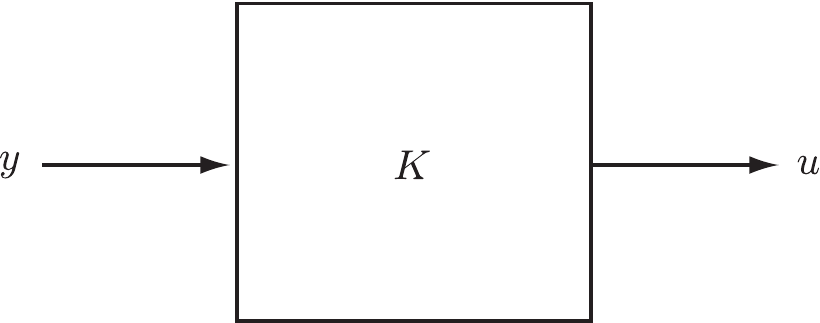}

\caption{A controller maps measurement records to control actions in a causal manner.}
\label{fig:controller1} 
\end{figure}

The control signal $u$ is determined by the controller $K$ using information contained in the observation signal $y$. The controller is to operate in real-time, so the controller is {\em causal}:
 \begin{quote}
 $u(t)$ depends on $y(s)$, $0 \leq s \leq t$
\end{quote}
In other words,  $u(t)$ is adapted to $\mathscr{Y}_t = \sigma \{ y(s), 0\leq s \leq t\}$, and we may write
 $u(t) = K_t( y(s), 0 \leq s \leq t)$, as in Figure \ref{fig:controller1}.

For a controller $K$ define the performance objective
\begin{equation}
J(K) = \mathbf{E}[ \int_0^T L(x(s), u(s)) ds + \Phi(x(T)) ]
\end{equation}
where (i)
  $L(x,u)$ and $\Phi(x)$ are suitably chosen cost functions reflecting the desired objective (e.g. regulation to a nominal state, say $0$), and (ii)
$\mathbf{E}$ denotes expectation with respect to the underlying probability distributions.

The optimal control problem is to minimize $J(K)$ over all admissible controllers $K$.
This is a {\em partially observed} stochastic optimal control problem: $J(K)$ is expressed in terms of the state $x$ which is not directly accessible. 
In order to solve this problem, we  now re-express $J(K)$ in terms of a new `state' that is accessible. Using basic properties of conditional expectation, we have
\begin{eqnarray}
J(K) &=&  \mathbf{E}[ \int_0^T L(x(s), u(s)) ds + \Phi(x(T)) ]
\\
&= &  \mathbf{E}[ \mathbf{E}[  \int_0^T L(x(s), u(s)) ds + \Phi(x(T))  \vert \mathscr{Y}_T ] ]
\\
&=&
\mathbf{E}[ \int_0^T  \tilde L(\pi_s, u(s))  ds +   \tilde \Phi(\pi_T) ]
\end{eqnarray}
where $\pi_t$ is the {\em conditional state}
\begin{equation}
\pi_t(\phi) = \mathbf{E}[ \phi(x(t)) \vert \mathscr{Y}_t ]
\end{equation}
and
\begin{equation}
\tilde L(\pi, u) = \pi( L(\cdot, u) ),  \ \tilde \Phi(\pi) = \pi(\Phi).
\end{equation}

The conditional state $\pi_t$ has the following relevant properties:
(i) 
$\pi_t$ is adapted to $\mathscr{Y}_t$, (ii)
the objective is expressed in terms of $\pi_t$, (iii)
$\pi_t$ is a Markov process (given $u$), with dynamics 
\begin{equation}
d \pi_t( \phi) = \pi_t( \mathcal{L}^{u(t)}(\phi)) dt + (\pi_t(\phi h) - \pi_t(\phi) \pi_t(h) ) (dy(t) - \pi_t(h) dt),
\end{equation}
the equation for nonlinear filtering \cite[Chapter 18]{RE82}. The conditional state $\pi_t$ is an example of an {\em information state}, \cite{KV86}.

An information state enables {\em dynamic programming} methods  to be used to solve the optimization problem.
Indeed, 
the {\em value function} is defined by
\begin{equation}
V(\pi,t)  = \inf_K \mathbf{E}_{\pi,t} [ \int_t^T \tilde L(\pi_s, u(s)) ds + \tilde \Phi( \pi_T  ) ],
\end{equation}
for which the corresponding
dynamic programming equation is
\begin{eqnarray}
\frac{\partial}{\partial t} V(\pi,t) + \inf_u \{ \tilde{\mathcal{L}}^u V(\pi ,t) + \tilde L(\pi,u) \} = 0 ,
\\
V(\pi,T) = \tilde \Phi(\pi) .
\nonumber
\end{eqnarray}
Here, $ \tilde{\mathcal{L}}^u$ is the generator for the process $\pi_t$.

If the dynamic programming equation has a suitably smooth solution, then the
optimal {\em feedback} control function
$$
\mathbf{u}^\star(\pi,t) = \displaystyle{\mathrm{argmin}_u} \{ \tilde{\mathcal{L}}^u V(\pi,t)  + \tilde L(\pi,u) \} 
$$
determines the optimal controller $K^\star$:
\begin{eqnarray}
d \pi_t( \phi) &=&  \pi_t( \mathcal{L}^{u(t)}(\phi)) dt + (\pi_t(\phi h) - \pi_t(\phi) \pi_t(h) ) (dy(t) - \pi_t(h) dt)
\label{eq:pi-rs-2}
\\
u(t) &=&  \mathbf{u}^\star(\pi_t,t) 
\label{eq:u-star-rs-2}
\end{eqnarray}

The optimal controller $K^\star$ has the well-known {\em separation structure}, where the dynamical part (the filtering equation (\ref{eq:pi-rs-2}) for the information state $\pi_t$) is concerned with estimation, and an optimal control part $\mathbf{u}^\star$ (\ref{eq:u-star-rs-2}), which determines control actions from the information state.
In the special case of Linear-Quadratic-Gaussian control, the conditional state is Gaussian, with conditional mean and covariance given by the 
Kalman filter,  while the optimal feedback $\mathbf{u}^\star$ is linear with the gain determined from the control LQR Riccati equation.

An alternative performance objective is the {\em risk-sensitive} performance objective \cite{J73,W81,BV85,JBE94}, defined for a controller $K$ by
\begin{equation}
J(K) = \mathbf{E}[ \exp(  \mu \{   \int_0^T L(x(s), u(s)) ds + \Phi(x(T)) \} ) ] ,
\label{eq:JK-rs-classical}
\end{equation}
where $\mu > 0$ is a risk parameter. 
Due to the exponential we cannot use the conditional state as we did above. Instead, we define an unnormalized  {\em risk-sensitive conditional state}
\begin{equation}
\sigma^\mu_t(\phi)= \mathbf{E}^0[ \exp(  \mu \{   \int_0^t L(x(s), u(s)) ds \})  \Lambda_t  \phi(x(t))  \vert \mathscr{Y}_t ]
\end{equation}
which includes the cost function $L(x,u)$. Here, the reference expectation is defined by
$$
\mathbf{E}^0[ \cdot ] = \mathbf{E}[ \cdot \Lambda_T^{-1}],
$$
where
$$
d \Lambda_t = \Lambda_t h(x(t)) dy(t), \ \ \Lambda_0=1.
$$
The risk-sensitive state $\sigma^\mu_t$ evolves according to
\begin{equation}
d \sigma^\mu_t( \phi) = \sigma^\mu_t( (\mathcal{L}^{u(t)}+ \mu L(\cdot, u(t) ) )\phi )    dt + 
\sigma^\mu_t(h) dy(t) .
\end{equation}
The performance objective may then be expressed as
\begin{equation}
J(K) = \mathbf{E}^0[  \sigma^\mu_T( e^{  \mu \Phi  } )].
\end{equation}
Thus $\sigma^\mu_t$ is an {\em information state} for the risk-sensitive optimal control problem, and we may use this quantity in dynamic programming.

The {\em value function} for the risk-sensitive problem is defined by
\begin{equation}
V^\mu(\sigma,t)  = \inf_K \mathbf{E}_{\sigma,t} [ \sigma^\mu_T( e^{  \mu \Phi  } )].
\end{equation}
The corresponding
dynamic programming equation is
\begin{eqnarray}
\frac{\partial}{\partial t} V^\mu(\sigma,t) + \displaystyle{\inf_u }\{ \tilde{\mathcal{L}}^{\mu, u} V^\mu(\sigma ,t)  \} = 0,
\\
V^\mu(\sigma,T) = \sigma( \exp( \mu \Phi )), 
\nonumber
\end{eqnarray}
where  $ \tilde{\mathcal{L}}^{\mu, u}$ is the generator for the process $\sigma^\mu_t$.
The optimal risk-sensitive  feedback  control function is
\begin{equation}
\mathbf{u}^{\mu,\star}(\sigma,t) = \displaystyle{ \mathrm{argmin}_u} \{ \tilde{\mathcal{L}}^{\mu, u} V(\sigma,t)  \} 
\end{equation}
and so the 
{\em optimal risk-sensitive controller} $K^\star$ is given by
\begin{eqnarray}
d \sigma^\mu_t( \phi) &=&  \sigma^\mu_t( (\mathcal{L}^{u(t)}+ \mu L(\cdot, u(t) ) )\phi )    dt + 
\sigma^\mu_t(h) dy(t)
\label{eq:rs-sigma-dyn-2}
\\
u(t) &=&  \mathbf{u}^{\mu, \star}(\sigma^\mu_t,t) .
\label{eq:rs-u-star-2}
\end{eqnarray}
Again, the optimal controller consists of a dynamical equation (\ref{eq:rs-sigma-dyn-2}) and a control function (\ref{eq:rs-u-star-2}), but estimation is not separated from control due to the cost term appearing in the filter  (\ref{eq:rs-sigma-dyn-2}).

\section{Quantum Measurement Feedback Optimal Control}
\label{sec:quantum}

\begin{figure}[htb]
\begin{center}
\includegraphics[scale=1.0]{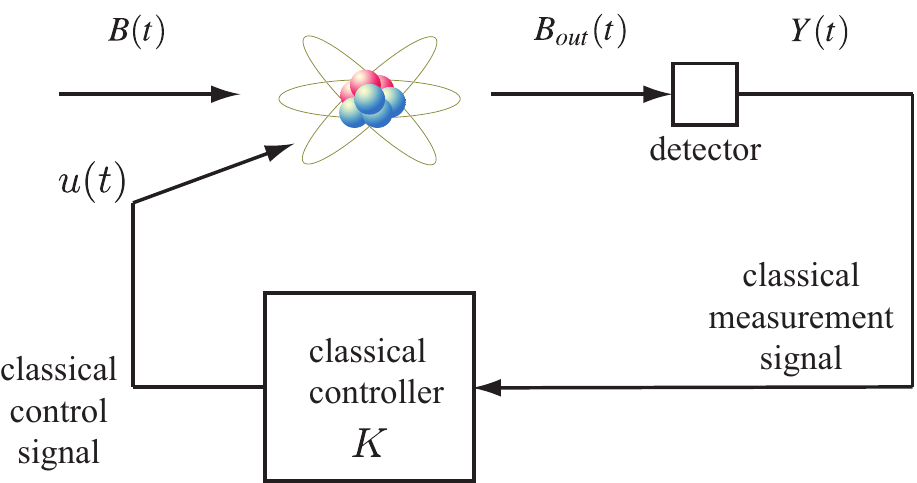}
\caption{An open quantum system controlled by a classical signal $u(t)$ and interacting with a quantum field. The output component of the field is continuously monitored producing an observation process $Y(t)$.}
\label{fig:mfb1}
\end{center}
\end{figure}

In this section we consider an extension of the optimal control results of the previous section to quantum systems. A schematic representation of the {\em measurement} feedback system is shown in Figure \ref{fig:mfb1}, where the classical system $K$ is the unknown controller to be determined.

 In what follows  we make use of {\em quantum stochastic differential equation} (QSDE) models for open quantum systems \cite{HP84,GC85,KRP92,GZ00}, and the theory of {\em quantum filtering} \cite{VPB92,VPB92a,HC93,BHJ07,WM10}.
 The {\em state} of an open quantum  system is specified by a state $\rho_0$ for the system (say atom) and a state for the environment, say the vacuum state $\Phi$ for the field. {\em Quantum expectation} $\mathbb{E}$ is given by $\mathbb{E}[ X \otimes F]= \mathrm{Tr}[ (\rho_0 \otimes \Phi) (X \otimes F)] = \mathrm{Tr}[ \rho_0 X] \mathrm{Tr}[\Phi F]$ for system operators $X$ and field operators $F$. Here, $\rho_0$ and $\Phi$ are density operators defined on the appropriate subspaces (system and environment).

In the QSDE framework for open quantum systems, dynamical evolution is determined by the 
{\em Schrodinger }  equation
\begin{equation}
dU(t) = \{L dB^\ast (t)- L^\ast dB(t) - (\frac{1}{2}
L^\ast L +iH(u) )dt \} U(t)
\end{equation}
for a unitary operator $U(t)$, 
where $B(t)$ is a {\em quantum Wiener process}.
System operators $X$ and output field $B_{out}(t)$ evolve according to the Heisenberg equations
\begin{eqnarray}
X(t)=j_{t}( X) =U^\ast( t) ( X\otimes I ) U( t) 
\\
 B_{out}(t) =U^\ast ( t) ( I \otimes  B(t) ) U( t) 
\end{eqnarray}
A standard measurement device (e.g. homodyne detector) is used to measure the following quadrature observable of the output field (see Figure \ref{fig:mfb1}):
\begin{equation}
Y(t) =  B_{out}(t) +  B_{out}^\ast(t) .
\end{equation}
For each $t$, the operator $Y(t)$ is self-adjoint, and for different times $t_1, t_2$, the operators $Y(t_1)$ and $Y(t_2)$ commute, and so by the spectral theorem \cite{BHJ07} $Y(t)$ is equivalent to a classical stochastic process (physically, a photocurrent measurement signal).

Using the quantum Ito rule, the system process $X(t)=j_t(X)$---a {\em quantum Markov process} (given $u$)---and output process $Y(t)$ are given by
\begin{eqnarray}
d j_t(X) &=&  j_t( \mathcal{L}^{u(t)}(X) )dt +  dB^\ast(t)   j_t(  [X,L]) + j_t( [ L^\ast, X] ) dB(t)
\\
dY(t) &=&  j_t( L+L^\ast) dt + dB(t) + dB^\ast(t)
\end{eqnarray}
where
\begin{equation}
\mathcal{L}^u(X) = -i [ X, H(u)] + \frac{1}{2} L^\ast [X,L] + \frac{1}{2} [L^\ast, X] L  .
\end{equation}

We denote by $\mathscr{Y}_t$ the {\em commutative} $\ast$-algebra of operators generated by the observation process $Y(s), 0 \leq s \leq t$. Since $j_t(X)$ commutes with all operators in $\mathscr{Y}_t$, the 
{\em quantum conditional expectation} 
\begin{equation}
\pi_t(X) = \mathbb{E}[ j_t(X) \vert \mathscr{Y}_t ]
\end{equation}
is well defined. 
The differential equation for $\pi_t(X)$ is called the
{\em quantum filter} \cite{VPB92,VPB92a,HC93,BHJ07}:
\begin{eqnarray}
d \pi_t(X) &=&  \pi_t( \mathcal{L}^{u(t)}(X)) dt 
\\
&& + ( \pi_t(XL + L^\ast X) - \pi_t(X) \pi_t(L+L^\ast)) (dY(t)-\pi_t(L+L^\ast)dt)
\end{eqnarray}

We now consider a quantum measurement feedback optimal control problem defined as follows. 
For a measurement feedback controller $K$ define the performance objective \cite{J05}\footnote{Earlier formulations of quantum measurement feedback optimal control problems were specified directly in terms of conditional states \cite{VPB83,DJ99}.}
\begin{equation}
J(K)= \mathbb{E}[ \int_0^T C_1(s) ds + C_2(T)  ],
\end{equation}
where 
(i)  $C_1(t) = j_t( C_1(u(t)))$ and $C_2(t)= j_t(C_2)$ are non-negative observables, and (ii) 
$\mathbb{E}$ denotes quantum expectation with respect to the underlying states for the system and field (vacuum). 
The  {\em measurement feedback quantum optimal control problem} is to minimize $J(K)$ over all measurement feedback controllers $K$, Figure \ref{fig:mfb1}.
Note that information about the system observables is not directly accessible, and so this is a partially observed optimal control problem.

Using standard properties of quantum conditional expectation, 
the performance objective  can be expressed in terms of the quantum conditional state $\pi_t$ as follows:
\begin{equation}
J(K) = \mathbb{E}[ \int_0^T \pi_s(C_1(u(s) ) ) ds + \pi_T( C_2)].
\end{equation}
Then dynamic program may be used to solve this problem, as in the classical case.
The
{\em optimal measurement feedback controller} has the separation form
\begin{eqnarray}
d \pi_t(X) &=&  \pi_t( \mathcal{L}^{u(t)}(X)) dt 
\\
&& + ( \pi_t(XL + L^\ast X) - \pi_t(X) \pi_t(L+L^\ast)) (dY(t)-\pi_t(L+L^\ast)dt) ,
\nonumber
\\
u(t) &=&  \mathbf{u}^\star(\pi_t,t) ,
\end{eqnarray}
where the feedback function $\mathbf{u}^\star$ is determined from the solution to a dynamic programming equation, see \cite{J05}.
Again the conditional state $\pi_t$  serves as an {\em information state}, this time for a quantum {\em measurement feedback} problem.

The risk-sensitive performance criterion (\ref{eq:JK-rs-classical}) may be extended to the present quantum context as follows, \cite{J05,WDDJ06}.
 Let $R(t)$ be defined by
\begin{equation}
\frac{dR(t)}{dt} = \frac{\mu}{2} C_1(t) R(t), \ \ R(0)=I.
\end{equation}
Then define the risk-sensitive cost to be
\begin{equation}
J^\mu(K) =  \mathbb{E}[ R^\ast(T) e^{\mu C_2(T) } R(T) ] .
\end{equation}
This definition accommodates in a natural way the observables in the running cost, which need not commute in general.
 
To solve this quantum risk-sensitive problem, we proceed as follows. 
Define $V(t)$ by
\begin{eqnarray}
d V(t) = \{ L dZ(t) +( - \frac{1}{2} L^\ast L 
-i H(u(t)) +\frac{\mu}{2} C_1(u(t)) 
)dt \}  V(t), \ \ V(0)=I, 
\nonumber
\end{eqnarray}
where $Z(t)=B(t)+B^\ast(t)$ (equivalent to a standard Wiener process with respect to the vacuum field state). The process
$V(t)$ commutes with all operators in the  commutative $\ast$-algebra $\mathscr{Z}_t$  generated   by $Z(s), 0 \leq s \leq t$.
We then have
\begin{equation}
J^\mu(K) = \mathbb{E}[ V^\ast(T) e^{\mu C_2} V(T)] .
\end{equation}
Next, define
an unnormalized  risk-sensitive conditional state
\begin{equation}
\sigma^\mu_t(X)  = U^\ast(t) \mathbb{E}[ V^\ast(t) X V(t) \vert \mathscr{Z}_t ] U(t)
\end{equation}
which evolves according to
\begin{eqnarray}
d \sigma^\mu_t(X) &=&  \sigma^\mu_t( (\mathcal{L}^{u(t)} + \mu C_1(u(t)) )X)) dt + \sigma^\mu(XL+L^\ast X) dY(t)
\end{eqnarray}
Then we have
\begin{equation}
J^\mu(K) = \mathbb{E}^0[ \sigma^\mu_T( e^{\mu C_2} ) ] ,
\end{equation}
and so  $\sigma^\mu_t$ serves as an {\em information state}, and the optimal risk-sensitive control problem may be solved using dynamic programming.

The
{\em optimal risk-sensitive measurement feedback controller} has the form
\begin{eqnarray}
d \sigma^\mu_t(X) &=&  \sigma^\mu_t( (\mathcal{L}^{u(t)} + \mu C_1(u(t)) )X)) dt + \sigma^\mu_t(L+L^\ast) dY(t)
\\
u(t) &=&  \mathbf{u}^{\mu\star}(\sigma^\mu_t,t) ,
\end{eqnarray}
where the feedback function $\mathbf{u}^{\mu\star}$ is determined from the solution to a dynamic programming equation, see \cite{J04,J05}.

The inclusion of a cost term in a quantum conditional state $\sigma^\mu_t$ appears to be new to physics, \cite{J04,J05,WDDJ06}.
This state depends on
(i)  information gained as the system evolves (knowledge), and
(ii) the objective of the closed loop feedback system (purpose).

\section{Coherent Quantum Feedback Control}
\label{sec:quantum-coherent}

\begin{figure}[h]
\begin{center}
\includegraphics{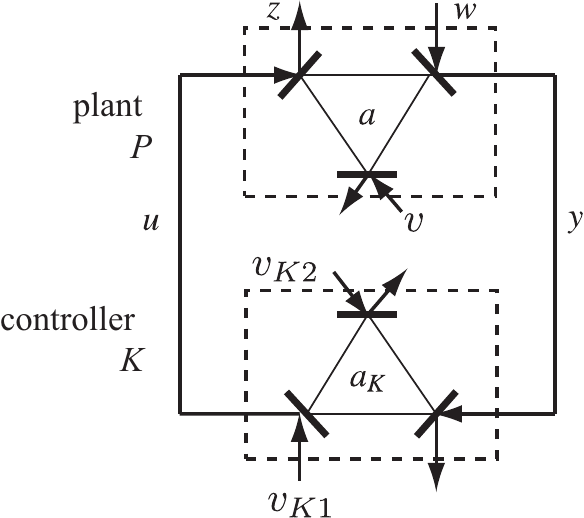}
\caption{Coherent feedback control example, showing plant $a$ and controller  $a_K$ cavity modes,
together with performance quantity $z$ and the \lq\lq{disturbance}\rq\rq \ input $w$.  The coherent signals $u$ and $y$ are used to transfer quantum information between the plant and the controller. The feedback system was designed to minimize the intensity of the  light at the output $z$ when an optical signal is applied at the input  $w$.}
\label{fig:hinfty1}
\end{center}
\end{figure}

An important challenge for control theory is to develop ways of {\em designing} signal-based coherent feedback systems in order to meet performance specifications, \cite{YK03a}, \cite{YK03b}, \cite{JNP08}, \cite{HM08}, \cite{NJP09}, \cite{GJ09}, 
\cite{NJD09}, \cite{JG10}.
While a detailed discussion of signal-based coherent feedback control design is beyond the scope of this article,  we briefly describe an example from \cite{JNP08}, \cite{HM08}. In this example, the plant is a cavity with three mirrors defining three field channels. The problem was to design a coherent feedback system to minimize the influence of one input channel $w$ on an output channel $z$, Figure \ref{fig:hinfty1}.  That is, if light is shone onto the mirror corresponding to the input  channel $w$, we would like the output channel $z$ to be dark. This is a simple example of robust control, where $z$ may be regarded as a performance quantity (to be minimized in magnitude), while $w$ plays the role of an external disturbance.
In \cite{JNP08}, it was shown how such problems could be solved systematically by extending methods from classical robust control theory, and importantly, taking into account the physical realization of the coherent controller as a quantum system. Indeed, the controller designed turned out to be another cavity, with mirror transmissivity parameters determined using mathematical methods.  This approach was validated by experiment \cite{HM08}.

Classical output feedback $H^\infty$ control problems can be solved through the use of a suitable information state, \cite{JB95,HJ99}. 
However, there is no known information state for the quantum coherent $H^\infty$ problem discussed above, and we now consider this matter more closely to see what concepts might be suitable for coherent feedback quantum control. 

Referring to Figure \ref{fig:hinfty1}, the plant $P$ and controller $K$ are connected by directional quantum signals $u$ and $y$ (beams of light). Such quantum signals may carry quantum information, and measurement need not be involved. 
The $H^\infty$ objective for the feedback network $P \wedge K$  is of the form
\begin{equation}
\mathbb{E}_{P \wedge K}[V(t)- V  -  \int_0^t S(r) dr ] \leq 0
\label{eq:hinfty-objective}
\end{equation}
where $V$ is a storage function  and $S$ is an observable representing the supply rate for the input signal $w$ and a performance variable $z$ (see \cite{JNP08,JG10} for general definitions of storage functions and supply rates). 
The storage function $V$ is a non-negative self-adjoint operator (observable). For example, for an optical cavity we may take $V=a^\ast a$, where $a$ and $a^\ast$ are respectively the annihilation and creation operators of the cavity mode (note that $V$ has spectrum $0,1,2,\ldots$, each value  corresponds to a possible number of quanta (photons) in the cavity).
A crucial difference between the fully quantum  coherent feedback and the measurement feedback situation discussed in Section \ref{sec:quantum} is that the algebra of operators $\mathscr{Y}_t$ generated by the plant output process $y(s), 0 \leq s \leq t$, is not commutative in general, and so a conditioning approach may not be feasible.

The controller $K$ shown in Figure \ref{fig:hinfty1} is an open quantum system that  involves additional quantum noise inputs $v_K$. These additional quantum noise terms are needed to ensure that $K$ is realizable as an open quantum system, and may be thought of as a \lq\lq{quantum randomization}\rq\rq (cf. classical randomized strategies).
The controller maps quantum signals as follows:
\begin{equation}
K : 
B_{K,in} = 
\left[ \begin{array}{c}
y
\\
v_{K1}
\\
v_{K2}
\end{array}
\right] \mapsto
B_{K,out} = 
\left[ \begin{array}{c}
z_{K1}
\\
u
\\
z_{K2}
\end{array}
\right]
\end{equation}

As an open system  not connected to the plant $P$, the controller $K$ has unitary dynamics given by a unitary operator $U_K(t)$ satisfying
\begin{equation}
dU_K(t) = \{L_K dB_{K,in}^\ast (t)- L_K^\dagger dB_{K,in}(t) - (\frac{1}{2}
L_K^\dagger L_K +iH_K )dt \} U_K(t), \ \ U_K(0)=I,
\label{eq:Ut-K}
\end{equation}
where $L_K=(L_{K0}, L_{K1}, L_{K2})^T$ and $H_K$ are the physical parameters determining the controller $K$ (an optical cavity, Figure \ref{fig:hinfty1}). This means that the input and output fields of the controller are related by
\begin{equation}
B_{K,out}(t)= U^\ast_K(t) B_{K,in}(t) U_{K}(t),
\end{equation}
 while the internal controller operators $X_K$ evolves according to $X_K(t) = j_{K,t}(X_K) = U^\ast_K(t) X_K U_K(t)$. In particular, the control field $u(t)$ is given by
 \begin{equation}
u(t) = U^\ast_K(t) v_{K1}(t) U_{K}(t),
\label{eq:u-coherent-1}
\end{equation}
or in differential form,
\begin{equation}
du(t) = j_{K,t}(L_{K1})dt + d v_{K1}(t)
\label{eq:u-coherent-2}
\end{equation}

Thus the controller $K$ is an open quantum system specified as follows:
\begin{equation}
K: \left\{   \begin{array}{l}
\mathrm{dynamics \ eq. \ (\ref{eq:Ut-K})}
\\
u(t) \ \mathrm{determined \ by \ (\ref{eq:u-coherent-1}) \ or \ (\ref{eq:u-coherent-2})} 
\end{array} \right.
\end{equation}
The controller $K$ has the property that it satisfies a performance objective of the form
\begin{equation}
\mathbb{E}_{K}[V_K(t) - V_K -  \int_0^t S_K(r) dr ] \leq 0,
\label{eq:hinfty-objective-2}
\end{equation}
and indeed a key step in classical approaches is such a reformulation of the original objective (\ref{eq:hinfty-objective}). The expression (\ref{eq:hinfty-objective-2})  does not (directly) involve the plant $P$, and $S_K$ is a suitable supply rate defined for the controller and the signals $u$ and $y$. The expectation is with respect to a state of the controller and not the plant. Furthermore, this property ensures that, when the controller $K$ is connected to the plant $P$, the feedback system $P \wedge K$ satisfies the objective (\ref{eq:hinfty-objective}).  In this way, the open system defining the controller $K$ serves as an {\em information system}, generalizing the concept of information state discussed in previous sections.

\section{Conclusion}
\label{sec:conclusion}

In this paper I have described how information states may be used to solve classical and quantum {\em measurement} feedback optimal control problems. Conditional expectation is a key mathematical tool that enables suitable information states to be defined. However, for fully quantum coherent feedback optimal control problems, the signals in the feedback loop are in general non-commutative quantum signals, and standard methods involving conditioning  are not applicable. Accordingly, 
I suggest that a concept of {\em information system} 
abstracting the notion of information state may provide a suitable means for approaching the solution of optimal fully quantum feedback control problems. Future work will be required to develop this idea further.

\bibliographystyle{plain}


\end{document}